\documentclass[journal, web, twocolumn]{IEEEtran}
\usepackage{mathtools} 
\usepackage{balance}

\usepackage{pifont} 

\usepackage{dsfont}

\usepackage{cite}
\usepackage{amsfonts}
\usepackage{amsmath}
\usepackage{mathrsfs}

\usepackage{graphicx}
\usepackage{textcomp}
\usepackage[table,xcdraw]{xcolor}
\usepackage{ragged2e}
\usepackage{bbm}
\usepackage{multirow}
\usepackage[linesnumbered, ruled]{algorithm2e}
\usepackage{algpseudocode}
\usepackage[caption=false, font=footnotesize]{subfig}
\usepackage[hang,flushmargin]{footmisc}
\usepackage{lipsum}
\makeatletter
\newcommand{\algorithmfootnote}[2][\footnotesize]{%
  \let\old@algocf@finish\@algocf@finish
  \def\@algocf@finish{\old@algocf@finish
    \leavevmode\rlap{\begin{minipage}{\linewidth}
    #1#2
    \end{minipage}}%
  }%
}
\usepackage[flushleft]{threeparttable}
\usepackage{etoolbox}

\appto\TPTnoteSettings{\footnotesize}

\newcommand{\thickhline}{%
    \noalign {\ifnum 0=`}\fi \hrule height 1pt
    \futurelet \reserved@a \@xhline
}
\newcolumntype{"}{@{\hskip\tabcolsep\vrule width 1pt\hskip\tabcolsep}}

\usepackage{array}
\newcolumntype{C}[1]{>{\centering\let\newline\\\arraybackslash\hspace{0pt}}m{#1}}
\def\BibTeX{{\rm B\kern-.05em{\sc i\kern-.025em b}\kern-.08em
    T\kern-.1667em\lower.7ex\hbox{E}\kern-.125emX}}

\makeatletter

\usepackage{hyperref}
\hypersetup{
    colorlinks=true,
    linkcolor=magenta,
    filecolor=blue,      
    urlcolor=cyan,
    citecolor=green,
}

\begin{document}

\title{A Lightweight CNN Model for Detecting Respiratory Diseases from Lung Auscultation Sounds using EMD-CWT-based Hybrid Scalogram}

\author{Samiul~Based~Shuvo\textsuperscript{1, ‡}, Shams~Nafisa~Ali\textsuperscript{1, ‡}, Soham~Irtiza~Swapnil\textsuperscript{1, ‡}, \\ 
         ~Taufiq~Hasan\textsuperscript{1},~\IEEEmembership{Member,~IEEE}~and~Mohammed~Imamul~Hassan~Bhuiyan\textsuperscript{2},~\IEEEmembership{Member,~IEEE}
\thanks{\textsuperscript{1}Samiul~Based~Shuvo, Shams~Nafisa~Ali, Soham~Irtiza~Swapnil and Taufiq~Hasan are with Department of Biomedical Engineering, Bangladesh University of Engineering and Technology (BUET), Dhaka-1205,
Bangladesh, Email: \{sbshuvo.bme.buet, snafisa.bme.buet, swapnil.buetbme\}@gmail.com, taufiq@bme.buet.ac.bd.}
\thanks{\textsuperscript{2}Mohammed~Imamul~Hassan~Bhuiyan is with Department of Electrical and Electronic Engineering, Bangladesh University of Engineering and Technology, Dhaka-1205, Bangladesh, Email: imamul@eee.buet.ac.bd}
\thanks{\textsuperscript{‡}These authors share first authorship on, and contributed equally to, this work.}
\thanks{Manuscript received August XX, 20XX; revised September XX, 20XX.}}

\markboth{IEEE JOURNAL OF BIOMEDICAL AND HEALTH INFORMATICS, VOL. XX, NO. XX, XXXX 2020}%
{Shuvo \MakeLowercase{\textit{et al.}}: IEEE JOURNAL OF BIOMEDICAL AND HEALTH INFORMATICS}

\maketitle

\begin{abstract}
Listening to lung sounds through auscultation is vital in examining the respiratory system for abnormalities. Automated analysis of lung auscultation sounds can be beneficial to the health systems in low-resource settings where there is a lack of skilled physicians. In this work, we propose a lightweight convolutional neural network (CNN) architecture to classify respiratory diseases using hybrid scalogram-based features of lung sounds. The hybrid scalogram features utilize the empirical mode decomposition (EMD) and continuous wavelet transform (CWT). The proposed scheme's performance is studied using a patient independent train-validation set from the publicly available ICBHI 2017 lung sound dataset. Employing the proposed framework, weighted accuracy scores of 99.20\% for ternary chronic classification and 99.05\% for six-class pathological classification are achieved, which outperform well-known and much larger VGG16 in terms of accuracy by 0.52\% and 1.77\% respectively. The proposed CNN model also outperforms other contemporary lightweight models while being computationally comparable.
\end{abstract}

\begin{IEEEkeywords}
Lung auscultation sound, respiratory disease detection, lightweight convolutional neural networks, empirical mode decomposition, continuous wavelet transform, scalogram.  
\end{IEEEkeywords}

%
\IEEEpeerreviewmaketitle


\section{Introduction}

\IEEEPARstart{L}{ung} diseases are the third largest cause of death in the world~\cite{s1}. According to the World Health Organization (WHO), the five major respiratory diseases~\cite{cruz2007global}, namely chronic obstructive pulmonary disease (COPD), tuberculosis, acute lower respiratory tract infection (LRTI), asthma, and lung cancer, cause the death of more than 3 million people each year worldwide~\cite{mathers2006projections, s2}. These respiratory diseases severely affect the overall healthcare system and adversely affect the lives of the general population. Prevention, early diagnosis and treatment are considered key factors for limiting the negative impact of these deadly diseases. 
\par 
Auscultation of the lung using a stethoscope is the traditional diagnostic method used by specialists and general practitioners for the initial investigation of the respiratory system. Although physicians use various other investigation strategies such as plethysmography, spirometry, and arterial blood gas analysis, lung sound auscultation remains as a vital tool for physicians due to its simplicity and low-cost~\cite{jayalakshmy2020scalogram}. 
The primary classification of these non-periodic and non-stationary sounds consists of two groups: normal (vesicular) and abnormal (adventitious)~\cite{abbas2010}. The first group is observed when there are no respiratory diseases, while the latter group indicates complications in the lungs or airways~\cite{reichert2008}. Crackle, wheeze, rhonchus, squawk, stridor, and pleural rub are the commonly known abnormal lung sounds. These anomalies can be differentiated from the normal lung sounds on the basis of frequency, pitch, energy, intensity, timbre, and musicality~\cite{sarkar2015, bohadana2014}. 
Therefore, lung sounds are of particular importance for recognizing specific respiratory diseases and assessing its chronic-nonchronic characteristics. 
However, the subtle differences between some of the adventitious lung sound classes can be a strenuous task even for a specialist and may introduce subjectivity in the diagnostic interpretation~\cite{bahoura}. In this scenario, artificial intelligence (AI)-powered algorithms can be of benefit in automatically interpreting lung sounds, especially in underdeveloped regions of the world, with a scarcity of skilled physicians. 
\par
In the past decade, a number of research approaches have been considered and evaluated for automatic identification of respiratory anomalies from lung auscultations sounds. Numerous feature extraction techniques including statistical features~\cite{palaniappan2013}, entropy-based features~\cite{zhang2009novel}, wavelet coefficients~\cite{bahoura2009pattern}, Mel Frequency Cepstral Coefficients (MFCC)~\cite{bahoura}, spectrograms~\cite{acharya2017feature}, scalograms~\cite{gautam2013wavelet}etc. have been adopted in conjunction with a diverse set of machine learning (ML) algorithms~\cite{bahoura, palaniappan2013, zhang2009novel, bahoura2009pattern, acharya2017feature, gautam2013wavelet, serbes2013pulmonary, iccer2014, jin2014new, bokov2016wheezing, mayorga2010acoustics, fenton1985automated, pasterkamp1997, dokur2009, rietveld1999}. 
\begin{figure*}[t]
    \centering
    \includegraphics[width=\linewidth]{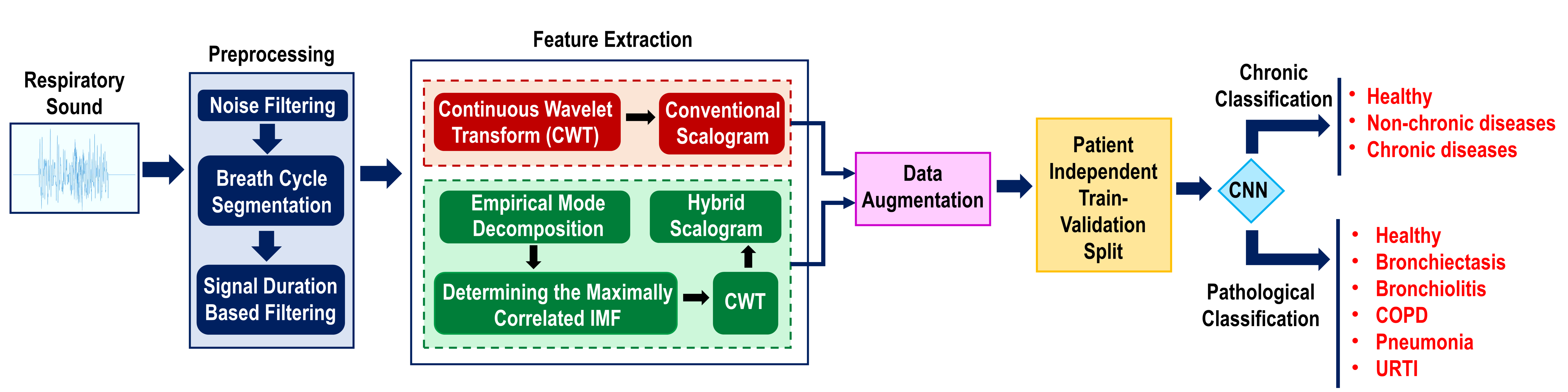}
    \caption{\textbf{A graphical overview of the proposed framework. After several generic preprocessing steps, the lung sound signals are converted into scalograms using both conventional and hybrid approaches. The resulting images are further augmented and fed into the proposed lightweight CNN model to carry out a two-way classification of respiratory diseases: (i) Chronic and (ii) Pathological.}}
    \label{f1}
\end{figure*}
\par 
With the advent of deep learning (DL), new developments have been made in recent times, demonstrating highly promising results in diversified applications, including biomedical engineering and clinical diagnostics~\cite{bozkurt2018, humayun2020towards, acharya2018, hershey2017cnn, debbal2004, meintjes2018}. With the ability of automatic feature learning, deep learning (DL) approaches are more generic and can mitigate the limitations of traditional ML-based methods. In the same vein, DL-based paradigms that are employed in the recent years for the identification of respiratory anomalies and pathologies from lung auscultation data have exhibited highly promising performance~\cite{jayalakshmy2020scalogram, minami2019, aykanat2017, demir2020, liu2019, bardou2018lung, acharya2020deep, perna2018cnn, perna2019deep, vbasu2020, pham2020robust, pham2020MoE, garcia2020vae}. However, for attaining proper functionality, the deep networks require to undergo an extensive training scheme with a large training dataset that subsequently calls for a considerable amount of time and the engagement of powerful computational resources. As a result, it becomes quite challenging to incorporate the deep learning frameworks in the currently available wearable devices and mobile platforms. In order to reduce the number of parameters of these networks, various methods have been investigated, including weight quantization~\cite{acharya2020deep}, lightweight networks~\cite{howard2017mobilenets}, and low precision computation~\cite{hubara2017quantized}.
\par
While constructing AI-assisted automated medical diagnosis frameworks, patient specificity in the train and validation dataset should be considered a salient factor to produce reliable results for unseen patient data, especially for chronic diseases \cite{kiranyaz2015, maheswari2008speaker}. Due to the available medical data's sparse nature, this factor is often neglected in the existing literature. The random adoption of 80\%-20\% or any other percentage of the train-validation split of the dataset corrupts most of the works with intra-patient dependency and ultimately the obtained results do not stand out to be consistent and generalized in case of a new patient~\cite{acharya2020deep}. Although this patient-independent division requires additional time and effort, the achieved results are more generalizable and represent the real-world scenarios.
\par 
In this work, a lightweight CNN architecture is proposed to perform respiratory disease classification (ternary chronic classification and six class pathology classification) utilizing the ICBHI 2017 scientific challenge respiratory sound database~\cite{data} while maintaining patient independent train-validation dataset splitting strategy. A hybrid approach for obtaining scalograms from respiratory sound signals is presented wherein continuous wavelet transform (CWT) is performed only on the maximally correlated intrinsic mode function (IMF) of the empirically decomposed (EMD) respiratory sound signals. The class discrimination capability of a hybrid scalogram is evaluated with respect to the CWT-based conventional scalogram. Subsequently, along with the proposed CNN model, complex CNN models such as VGG16~\cite{vgg}, AlexNet~\cite{alexnet} and several contemporary lightweight architectures including MobileNet V2~\cite{mobilenetv2}, NASNet~\cite{nasnet} and ShuffleNet V2~\cite{shufflenet} are used for classifying the scalogram images to detect respiratory diseases in different categories. A comparative study among the proposed CNN model and the others is presented in terms of detection performance and being a lightweight network.
\par 
The rest of the paper is organized as follows. Previous studies related to lung sound classification using different ML-based approaches are discussed in Section~\ref{mt2}. Section~\ref{mt3} describes the dataset, feature extraction process, and the proposed lightweight CNN model. The experimental setup and results are discussed in Section~\ref{mt4}. The performance of the proposed method is compared with other works in Section~\ref{mt5}. Finally, the concluding remarks are provided in Section~\ref{mt7}.

\section{Related Works} \label{mt2}
Many research works employing machine learning, and deep learning have been reported on developing automated systems for respiratory sound classification.  However, the majority of the works have focused on respiratory anomaly prediction, basically classifying the lung sounds as wheeze, crackles~\cite{bahoura, palaniappan2013, zhang2009novel, bahoura2009pattern, acharya2017feature, gautam2013wavelet, serbes2013pulmonary, iccer2014, jin2014new, bokov2016wheezing, mayorga2010acoustics, fenton1985automated, pasterkamp1997, dokur2009, rietveld1999, minami2019, aykanat2017, demir2020, liu2019, bardou2018lung, acharya2020deep} rather than directly predicting respiratory diseases from lung auscultation recordings. The few works geared towards pathology classification are very recent and mostly involve elaborate processing or dedicated CNN and RNN frameworks due to the inherent complexity of the signal ~\cite{perna2018cnn, perna2019deep, vbasu2020, pham2020robust, pham2020MoE, garcia2020vae}. However, at pathology-level, so far, the classification task has been investigated at three different resolutions; the binary classification (healthy, pathological)~\cite{perna2018cnn, perna2019deep}, the ternary chronic classification (healthy, chronic disease, non-chronic disease)~\cite{perna2019deep, garcia2020vae} and multi-class distinct disease classification~\cite{vbasu2020, garcia2020vae}. Among the diseases, Upper and Lower Respiratory Tract Infection (URTI and LRTI), bronchiolitis and pneumonia have been included in the non-chronic disease class while COPD, asthma and bronchiectasis have been combined to form the chronic class~\cite{perna2019deep}. 
\par 
In~\cite{perna2018cnn}, a novel CNN based ternary classification approach has been implemented and performed considerably well with 82\% accuracy and 88\% ICBHI score. Later, the same authors proposed a Mel-Frequency Cepstral Coefficient (MFCC) and Long Short-term Memory (LSTM) based framework capable of conducting both binary and ternary classification of respiratory diseases~\cite{perna2019deep} which demonstrated excellent performance with 99\% and 98\% accuracy, respectively. Another work involving complex RNN architecture and extensive preprocessing has reported accuracy of 95.67\%±0.77\% in predicting six class pathology-driven diseases~\cite{vbasu2020}. However, by employing a CRNN network with a CNN-Mixture-of-Experts (MoE) baseline to learn both spatial and time-sequential features from the spectrograms, recent work has achieved a specificity of 83\% and a sensitivity of 96\% in ternary respiratory disease classification~\cite{pham2020robust}. For binary classification, the same work has reported specificity and sensitivity of 83\% and 99\%, respectively. As an extension of~\cite{pham2020robust}, a separate study involving the robust Teacher-Student learning schemes with knowledge distillation has been conducted, which resulted in a substantially reduced specificity while maintaining the sensitivity~\cite{pham2020MoE}.
\par 
Since the existing heavily imbalanced datasets of lung auscultations further exacerbate the task of respiratory disease classification, a contemporary study has dealt with this issue by experimenting with several data augmentation techniques, such as SMOTE, Adaptive Synthetic Sampling Method (ADASYN) and Variational autoencoder (VAE)~\cite{garcia2020vae}. Among the methods, the VAE-based Mel-spectrogram augmentation strategy, in conjunction with a CNN model, has achieved the best results with 98.5\% sensitivity and 99.0\% specificity in ternary chronic classification. The strategy has also exhibited an equally sophisticated performance with 98.8\% sensitivity and 98.6\% specificity in the case of six class respiratory disease classification~\cite{garcia2020vae}. 
\par 
Although the scope of DL-based frameworks with a spectrogram-based feature extraction strategy has been investigated in several works for direct classification of respiratory diseases from lung auscultations~\cite{pham2020robust, pham2020MoE, garcia2020vae}, to the best of the knowledge of the authors, scalogram based approaches have not been explored in this domain. Additionally, no dedicated lightweight, efficient CNN framework has been developed and investigated for the respiratory disease classification task. Furthermore, none of the studies consider the issue of intra-patient dependency in the train-validation split. Inspired by all of these factors, a scalogram based approach in conjunction with a lightweight CNN is proposed in this paper for the prediction of respiratory diseases from lung auscultations, maintaining patient independence. The proposed framework is schematically represented in Fig.~\ref{f1}.


\section{Materials and Methods}    \label{mt3}
\subsection {ICBHI 2017 Dataset}
ICBHI (International Conference on Biomedical Health Informatics) 2017 database is a publicly available benchmark dataset of lung auscultations~\cite{data}. It is collected by two independent research teams of Portugal and Greece. The dataset contains 5.5 hours of audio recordings sampled at different frequencies (4 kHz, 10 kHz, and 44.1 kHz), ranging from 10s to 90s, in 920 audio samples of 126 subjects from different anatomical positions with heterogeneous equipment~\cite{data2}.
\par
The samples are professionally annotated considering two schemes: 1. according to the corresponding patient’s pathological condition, i.e. healthy and seven distinct disease classes, namely Pneumonia, Bronchiectasis, COPD, URTI, LRTI, Bronchiolitis, Asthma and 2. according to the presence of respiratory anomalies, i.e. crackles and wheezes in each respiratory cycle. Further details about the dataset and data collection methods can be found in~\cite{data2}.

\subsection {Data Prepossessing}
\subsubsection{Noise filtering}
Since 50 Hz to 2500 Hz is the acknowledged frequency range of the lung auscultation signals~\cite{reichert2008}, the recorded audio signals are filtered with a 6th order Butterworth bandpass filter, thus retaining 50 Hz to 2500 Hz frequency components. Subsequently, all the sample signals are resampled to 22050 Hz for ensuring consistency and normalized to the range [-1,1] for attaining device homogeneity.
\subsubsection{Segmentation of the sound data}
Each of the audio recordings is segmented according to the annotated respiratory cycle timing with a 6s duration each. Samples with a minimum respiratory cycle duration of 3s are taken into account to obtain useful respiratory sound information~\cite{pham2020robust}. Post performing this procedure, 2 of the disease classes, namely Asthma and LRTI, are found to have inadequate segmented samples for meaningful feature extraction and therefore, these two classes are not considered in our study. After these procedures, lung auscultation sounds from 87 out of 120 independent patients are usable. Table~\ref{t2} represents data distribution at several levels of processing corresponding to the disease classes considered for this study. 

\begin{table}[h!]
\centering
\caption{Distribution of Data at Different Processing Levels Corresponding to the Disease Classes}
\label{t2}
\begin{tabular}{|p{1.5cm}|C{1.25cm}|C{1.5cm}|C{1cm}|C{1.5cm}|}
    \hline
    \textbf{Disease Name}&\textbf{No. of unsegmented sound file }&\textbf{	No. of Segmented and Filtered Sample}&\textbf{	No. of Unique Patient}&	\textbf{No. of Generalized Augmented Image}\\
    \hline
    Pneumonia&37&	41&	3&	164\\
    \hline
    Bronchiectasis&	16&55&	6&	220\\
	\hline
	COPD&793&	1,963&	51&	1963\\
	\hline
    Healthy &35&	42&	13&	168\\
	\hline
	URTI&	23&21	&8&	84\\
	\hline
    Bronchiolitis&13	&65	&6&	260\\
	\hline
     Total &917	&2187&	87	&2859\\
     \hline
\end{tabular}
\end{table} 

\subsection{Feature extraction}
\subsubsection{Empirical Mode Decomposition (EMD)}
EMD is a powerful self-adaptive signal decomposition method especially in the time scale and energy distribution aspects and highly suitable for analysis and processing of non-linear and non-stationary signals such as lung sounds and heart sounds~\cite{ibtehaz2019}. It decomposes a given signal \textit{x(t)} into a finite set (N) of intrinsic mode functions, \textit{IMF\textsubscript{1}(t), IMF\textsubscript{2}(t), . . . . , IMF\textsubscript{N}(t)}, depending on the local characteristic time scale of the signal, with a view to expressing the original signal as the sum of all its IMF plus a final trend either monotonic or constant called residue, \textit{r(t)} a: \( x(t)\ = \sum_{i=1}^{N}{{IMF}_i(t)\ +~r(t)}\)~\cite{altuve2020fundamental}. An IMF is a simple oscillatory function with the equal number of extrema and zero crossings and its envelopes must be symmetrical with respect to zero. Thus, the EMD detrends a signal and elicits underlying spectral patterns~\cite{ibtehaz2019}.

\subsubsection{Continuous Wavelet Transform (CWT)}
Wavelet transform is defined as a signal processing method that can decompose a signal into an orthonormal wavelet basis or into a set of independent frequency channels~\cite{gautam2013wavelet, debbal2004}. Using a basis function, i.e. the mother wavelet \textit{g(t)}, and its scaled and dilated versions, the Continuous Wavelet Transform (CWT) can be used to decompose a finite-energy signal, \textit{x(t)} as~\cite{meintjes2018}: 
\begin{equation}
  Z(a,b) = \int {x(t)*g(t)(\frac{t-a}{b})}
    \end{equation}
where b denotes time location and a is scale factor.
Larger scale values reveal low-frequency information while the smaller scale values reveal high-frequency information~\cite{debbal2004}. The squared-modulus of the CWT coefficients Z is known as the scalogram~\cite{gautam2013wavelet}.

\subsection {Scalogram Representations}

\subsubsection{Conventional Scalogram}
\begin{figure}[b]
    \centering
    \includegraphics[width=3.5 in]{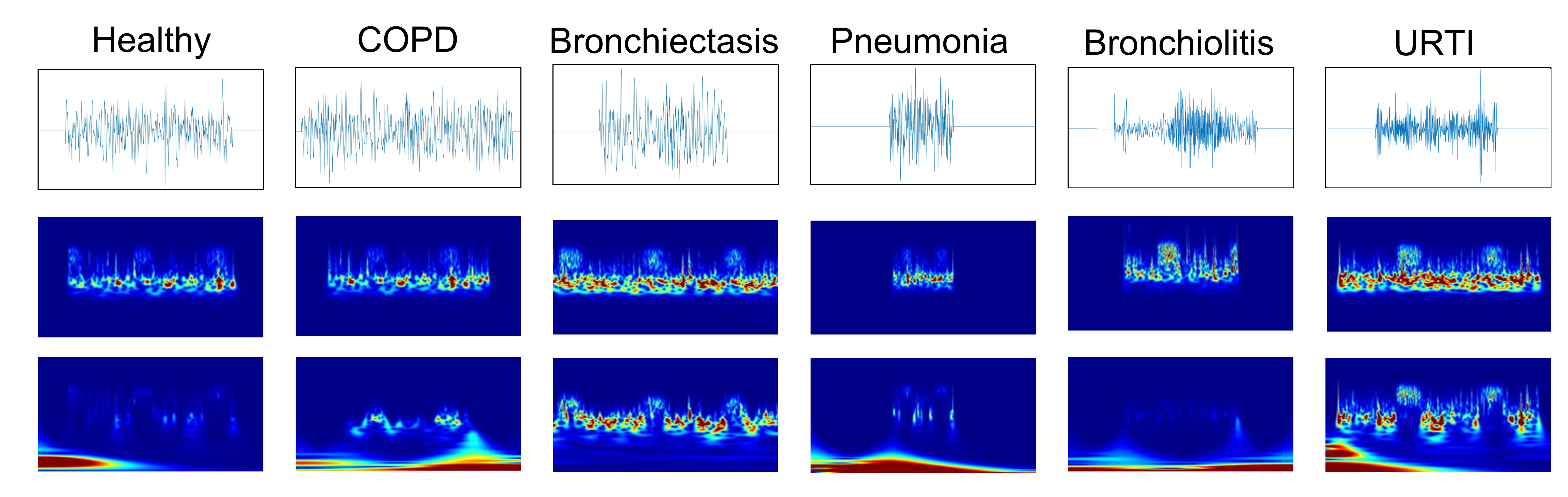}
    \caption{\textbf{Scalograms of the lung auscultation sounds for 6 disease classes; lung sound recordings (1st row), conventional scalogram (2nd row) and scalogram using the proposed hybrid approach (3rd row).}}
    \label{f2}
\end{figure}

Scalogram is defined as the time-frequency representation of a signal that depicts the obtained energy density using CWT~\cite{jayalakshmy2020scalogram, ren2018}. The segmented and filtered lung sound samples are decomposed into corresponding wavelet coefficients in MATLAB 2020a by using Morse analytic wavelet. Scalogram plots are generated with a resolution of 224 × 224 using these coefficients. Fig.~\ref{f2} shows the scalograms of lung sounds in different disease categories.

\subsubsection{Hybrid Approach for Scalogram} 
For each segmented and filtered sample under each pathological class, 9 IMFs are generated using the EMD function in MATLAB 2020a. Based on the cross-correlation between the source signal and the IMFs, the most physically significant IMF output with the highest correlation coefficient is determined~\cite{fontugne2012EMD1, boutana2014EMD2}. Subsequently, the squared-modulus of the CWT of the corresponding IMF is calculated to obtain the scalogram.  
\par
\begin{figure}[h!]
    \centering
    \includegraphics[width=3.5 in]{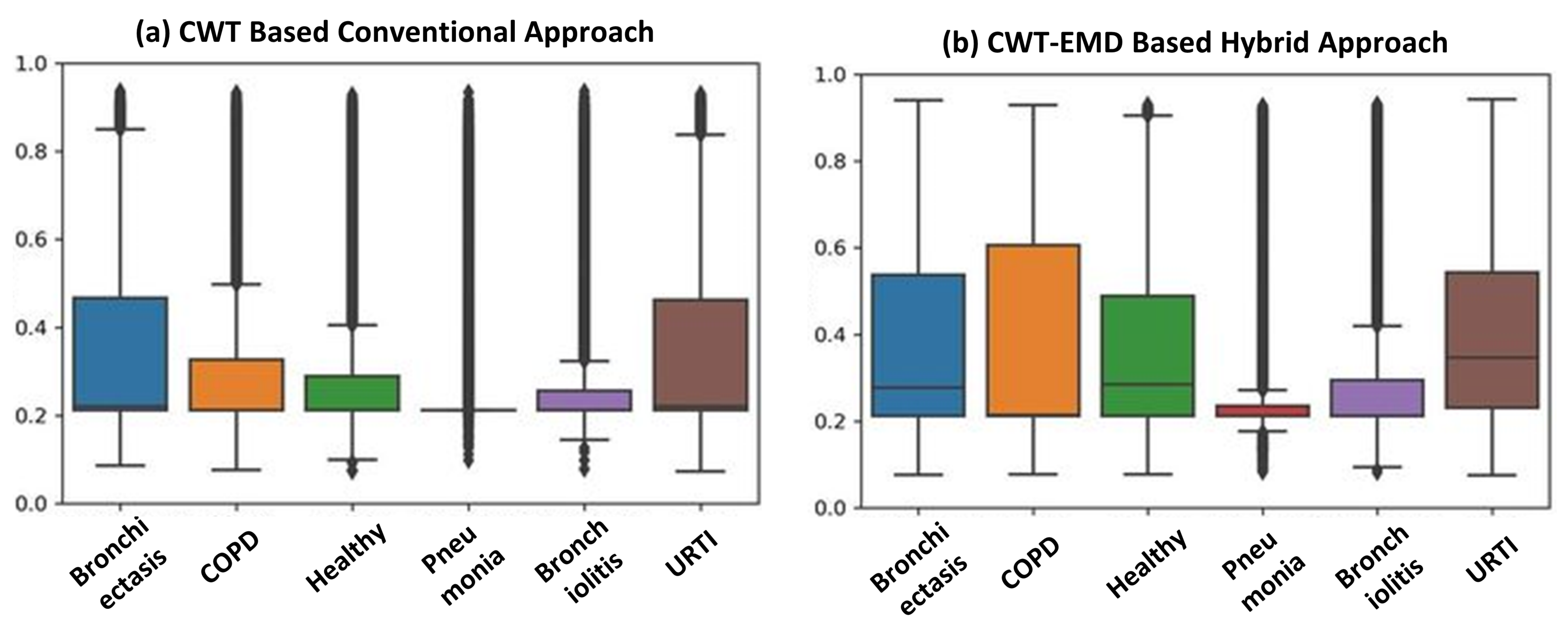}
    \caption{\textbf{Box plot. (a) Scalogram using the conventional CWT approach; (b) Scalogram using the hybrid approach.}}
    \label{f3}
\end{figure}

The diverse frequency bands varying from the maximum to the minimum range give the IMFs the capability to extract the temporal and spectral information~\cite{altuve2020fundamental} effectively. Hence, when this IMF based scheme is combined with CWT-oriented scalogram representation, the newly formed hybrid scalograms can demonstrate more discriminative and significant features. Thus, it has the potential to provide better classification performance by a CNN model. The box plots of the scalograms of lung sounds for various respiratory diseases are shown in Fig.~\ref{f3}. The distinction among the plots is more evident when using the hybrid approach than those of the conventional scalograms obtained using only CWT. 
\par
It should be mentioned that the proposed scalogram is distinctly different from that of ~\cite{jayalakshmy2020scalogram,boutana2014EMD2} in that the CWT modulus is computed from the maximally correlated IMFs and thus, providing a better representation of the underlying information. Note that the works of  ~\cite{jayalakshmy2020scalogram} and~\cite{boutana2014EMD2} are on detecting respiratory anomalies such as crackle and wheeze, and analysis and segmentation of heart sounds, respectively, whereas our objective is to detect respiratory diseases from the lung auscultations.

\subsection{Augmentation}
The ICBHI 2017 dataset is highly imbalanced, with around 86\% of the data belonging to COPD. Image augmentation using different color mapping schemes is employed to oversample the less represented classes and address the data imbalance issue~\cite{shih2019}. Colormaps are three-column arrays containing RGB triplets where each row defines a distinct color. Scalogram representation using different color maps helps generalize the produced images.
\par 
From each of the audio samples of the less represented data classes, four scalograms are generated for each segmented sample using four different color mapping schemes: Parula, HSV, Jet, and Hot, which are available in MATLAB 2020a while for the most represented class, COPD, only one image is produced from each audio sample. Nevertheless, for ensuring generalization and homogeneity of the augmented data, all four-color mapping schemes are randomly utilized for COPD. A summary of segmented audio files and final augmented scalogram images with corresponding diseases classes are presented in Table~\ref{t2}.
    
\section{Proposed lightweight CNN architecture}    \label{mt4}
CNN has become a popular approach for classifying image data, and recently there have been several works using CNN on classifying images produced from sounds~\cite{jayalakshmy2020scalogram, minami2019, demir2020}. However, due to memory constraints, a regular deep CNN model is computationally expensive with its large number of learnable parameters and arithmetic operations. Thus, it is not suitable for embedded devices as they cannot afford the processing complexity and storage space for parameters and weight values of filters~\cite{acharya2020deep}. 
Cloud computing methodology requires a higher RAM for this computationally intensive training and hence are outsourced~\cite{nikouei2018}. For this reason, Lightweight CNN models are gaining popularity among researchers for their faster performance and compact size without compromising the much-needed accuracy performance compared to the well-known deep learning networks~\cite{lim2018}. 
\par
\begin{figure*}[t]
    \centering
    \includegraphics[width =7in]{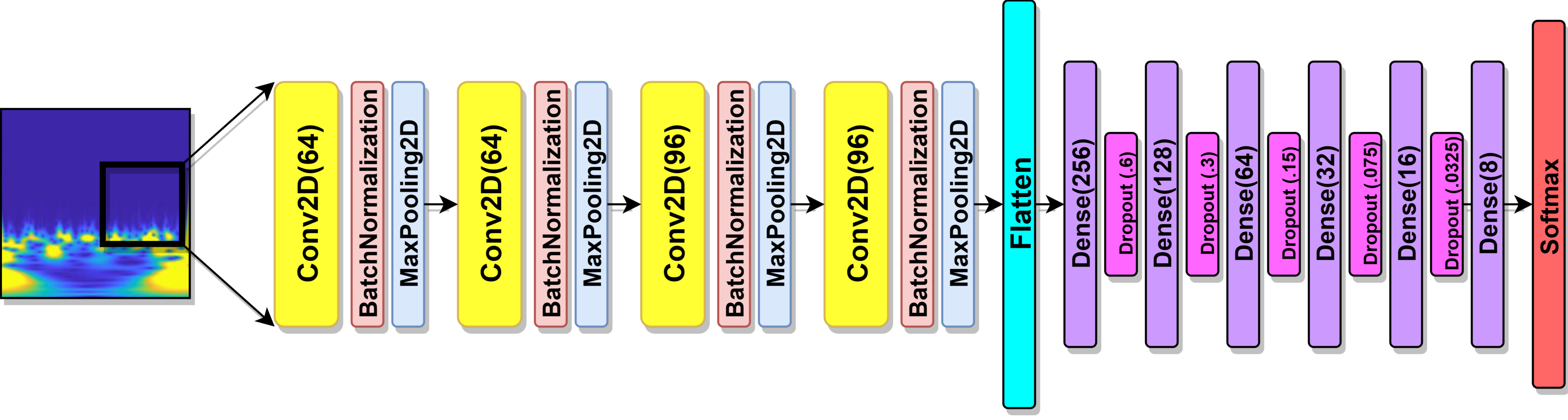}
    \caption{\textbf{The detailed architecture of the proposed lightweight CNN model.}}
    \label{f5}
\end{figure*}

The architecture of the proposed CNN model consists of an input layer corresponding to the 3-channel input of 224×224 images. The architecture of the proposed model is illustrated in Fig.~\ref{f5}.
\par 
The 1st convolutional layer uses 64 output filters with a 5×5-pixel size kernel followed by a 2×2-pixel max-pooling layer. Three additional convolutional layers are stacked over the first layer, each having a 3×3-pixel size kernel with 64, 96, and 96 filters sequentially and corresponding batch-normalization and max-pooling layers with 2×2 pooling window. Outputs from all these layers are flattened and connected with five pairs of FC and dropout layers, followed by a SoftMax output layer with probability nodes for each class. ReLU activation layer is applied within convolution calculation and fully connected layers and employed to introduce nonlinearity within the calculation and reduce the time for convergence. It does not get activated for any negative value. Max pooling is used after the ReLU activation; it reduces the spatial dimensionality of the extracted feature maps, extracts the most important features, and is unaffected from locational bias~\cite{bardou2018lung}.  To overcome the problem of more diverse data variance, the Batch Normalization layer with every convolution layer normalizes the extracted feature. It gives the network a representative power with a small number of parameters and faster training capability by reducing the variance.

\section{Experimental Results}    \label{mt5}
\subsection{Evaluation Criteria}
The augmented image sets are divided into 80\% training and 20\% validating parts for training and fine-tuning the model hyperparameters. Patient uniqueness, a critical aspect in the real-world applications, is maintained while dividing into training and validation parts as speaker dependency results in biased accuracy~\cite{maheswari2008speaker}. 
\par 
The classifier models' performance is evaluated based on the well-known evaluation matrices, namely, accuracy, recall (sensitivity), precision, and F1-score. Additionally, specificity and ICBHI-score~\cite{perna2019deep, data2}, a dedicated metric involving both sensitivity and specificity to assess the performance of the frameworks using the ICBHI dataset, is used to evaluate the performance of our method. 

\begin{table*}[h]
\centering
\caption{{Summary of the Classification Performance \break The \textcolor{red}{red} and \textcolor{blue}{blue} marked values represent the highest accuracy obtained with our proposed CNN model and VGG16}}

\label{tab3}
\begin{tabular}{|p{1cm}|p{.5cm}|p{.60cm}|p{.5cm}|p{.52cm}|p{.5cm}|p{.60cm}|p{.5cm}|p{.52cm}|p{.5cm}|p{.60cm}|p{.5cm}|p{.52cm}|p{.5cm}|p{.60cm}|p{.5cm}|p{.52cm}|}
\hline
\multirow{3}{*}{\textbf{Network}} & \multicolumn{8}{c|}{\textbf{Chronic Classification}}                                                                                                          & \multicolumn{8}{c|}{\textbf{Pathological Classification}}                                 \\ \cline{2-17} 
                                  & \multicolumn{4}{c|}{\textbf{Scalogram using CWT}}                                     & \multicolumn{4}{c|}{\textbf{Scalogram using EMD and CWT}}                                   & \multicolumn{4}{c|}{\textbf{Scalogram using CWT}}   &
                                  \multicolumn{4}{c|}{\textbf{Scalogram using EMD and CWT}}   \\ \cline{2-17} 
                                  & \textbf{Prec.}  & \textbf{Recall} & \textbf{Acc.}  & \textbf{F1}    & \textbf{Prec.}  & \textbf{Recall} & \textbf{Acc.}  & \textbf{F1}    & \textbf{Prec.}  & \textbf{Recall} & \textbf{Acc.}  & \textbf{F1} &
                            \textbf{Prec.}  & \textbf{Recall} & \textbf{Acc.}  & \textbf{F1}\\ \hline
 \textcolor{red}{Proposed model} & 95.00& 91.00& 90.58        & 92.00          & 99.25         & 99.20        &  \textcolor{red}{99.20}        & 99.22        & 87.00         & 86.00         & 86.31          & 86.00 & 99.12          &  99.05         &  \textcolor{red}{99.05}          & 98.96          \\ \hline
 \textcolor{blue}{VGG16}   & 92.00        & 89.00        & 88.58       & 90.00          & 99.03          & 98.69  &  \textcolor{blue}{98.69}         & 99.11         & 85.00          & 85.00        & 85.80 & 85.00 & 97.80      & 97.32        &  \textcolor{blue}{97.32}          & 97.01          \\ \hline                   

\end{tabular}
\end{table*}
\begin{table*}[t]
\caption{Comparison  of the Proposed Framework with Existing Works Using the ICBHI 2017 Dataset}
\label{t7}
\begin{center}
\begin{tabular}{|c|c|c|c|c|c|c|}
    \hline
  \textbf{ Processing }& \textbf{	Type of Training Network} &	 \textbf{Number of Prediction Classes} & \textbf{	Acc.}& \textbf{	Spec.}	 & \textbf{Sen.} &	 \textbf{ICBHI Score} \\    \hline
\multirow{3}{*}{Gammatone~\break
Spectrogram~\cite{pham2020robust, pham2020MoE}}&C-RNN &\multirow{3}{*}{3 (Healthy, Chronic, Non-chronic)}&-&	0.57&	0.94&	0.76\\  
&CNN-MoE& & -&	0.86&	0.96&	0.91\\
&Ensemble& & -&	0.71&	0.95&	0.83\\\hline
\multirow{2}{*}{MFCC~\cite{perna2019deep}} &CNN &\multirow{2}{*}{3 (Healthy, Chronic, Non-chronic)}&0.82&	0.76
&	0.89&	0.83\\  
&LSTM&&0.98&	0.82&	0.98&	0.90\\\hline
MFCC~\cite{vbasu2020}&RNN&	6 classes (excluding Asthma, LRTI)&	0.9567&	-&	0.9567&	-\\    \hline
\multirow{2}{*}{Mel-spectrogram-
\break VAE~\cite{garcia2020vae}} &\multirow{2}{*}{CNN}&3 (Healthy, Chronic, Non-chronic)	&	0.99	&0.990&	0.985&	0.988\\
&&6 classes (excluding Asthma, LRTI)&	0.99&	0.986&	0.988&	0.987
   \\ \hline
\multirow{2}{*}{\bf{Hybrid Scalogram}
\break \bf(Proposed)}&\multirow{2}{*}{Lightweight CNN}&3 (Healthy, Chronic, Non-chronic)& \bf{0.99}&	1.00&	0.99&	0.995\\
&&6 classes (excluding Asthma, LRTI)&	 \bf{0.99}&	1.00&	0.99&	0.995\\
\hline
\end{tabular}
\end{center}
\end{table*}

\subsection{Experimental Setup}
The proposed CNN model is constructed using Keras and TensorFlow backend, and trained using NVidia K80 GPUs provided by Kaggle notebooks. The mini-batch training scheme is employed while feeding the image data into a model for tackling the class imbalance issue. This technique performs by oversampling the scarce classes while randomly undersampling a majority class. This strategy ensures that the CNN model takes an equal number of samples from each class during each of the training epochs and thereby forms a balanced training set~\cite{humayun2020towards}. 
\par 
The adaptive learning rate optimizer (Adam) with the learning rate of 0.00001 is used for compiling the model. The batch size needs to be a multipler of 6 since an equal number of samples from each of the 3 and 6 data classes are taken in each training and validation batch~\cite{humayun2020towards}. In this study, batch size 6 has been taken for training and validation of both the classification schemes.
\par
As stated earlier, both the ternary chronic classification (chronic, non-chronic, healthy) and six class (Bronchiectasis, Bronchiolitis, COPD, Healthy, Pneumonia, and URTI) pathological classification are carried out in this work. The classification performance of the proposed CNN model is compared with that of VGG16, a well-known CNN architecture for image classification~\cite{vgg} in both of the classification schemes. It should be noted that the experiments are performed using both the convention CWT-based scalogram and hybrid scalogram images. In addition, the performance of our proposed CNN model is compared with a number of well-known DL architectures such as VGG16 and AlexNet and several lightweight networks in terms of computational complexity and accuracy.
\begin{figure}[b]
    \centering
    \includegraphics[width=3.5 in]{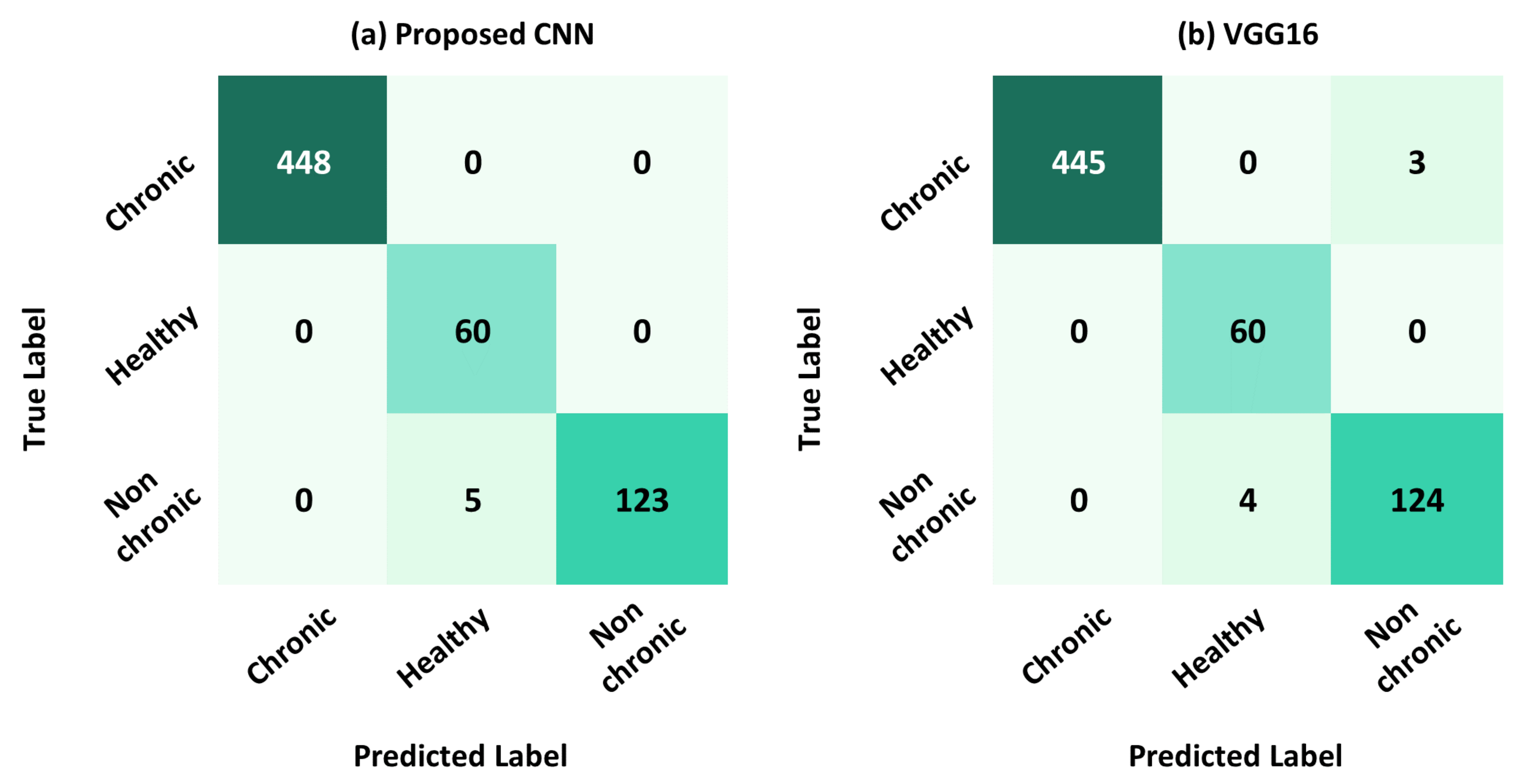}
    \caption{\textbf{Confusion matrices for the best results obtained in ternary chronic classification. (a) Proposed CNN model with batch size 6; (b) VGG16 with batch size 6.}}
    \label{f6}
\end{figure}

\subsection{Classification Performance of the Proposed Framework}
\subsubsection {Chronic Classification}
From Table~\ref{tab3}, it can be seen that using the hybrid scalogram method in conjunction with the proposed CNN model classifier shows the best accuracy, 99.21\%. However, the corresponding accuracy obtained by using VGG16 is quite close (98.89\%). Despite being a heavy model, the comparatively lower accuracy of VGG16 can be attributed to the over-fitting issue due to the limited number of images in different classes. When comparing conventional CWT scalogram to the proposed hybrid scalogram, considerable improvement in accuracy is evident for the latter using VGG16 and our proposed CNN model (9.5\%-11.4\%). The corresponding confusion matrices for both the models' best results are illustrated in Fig.~\ref{f6}. The results depict that the proposed method is better in ternary chronic classification than the VGG16.

\subsubsection {Pathological Classification} 
\begin{figure}[b]
    \centering
    \includegraphics[width=3.5 in]{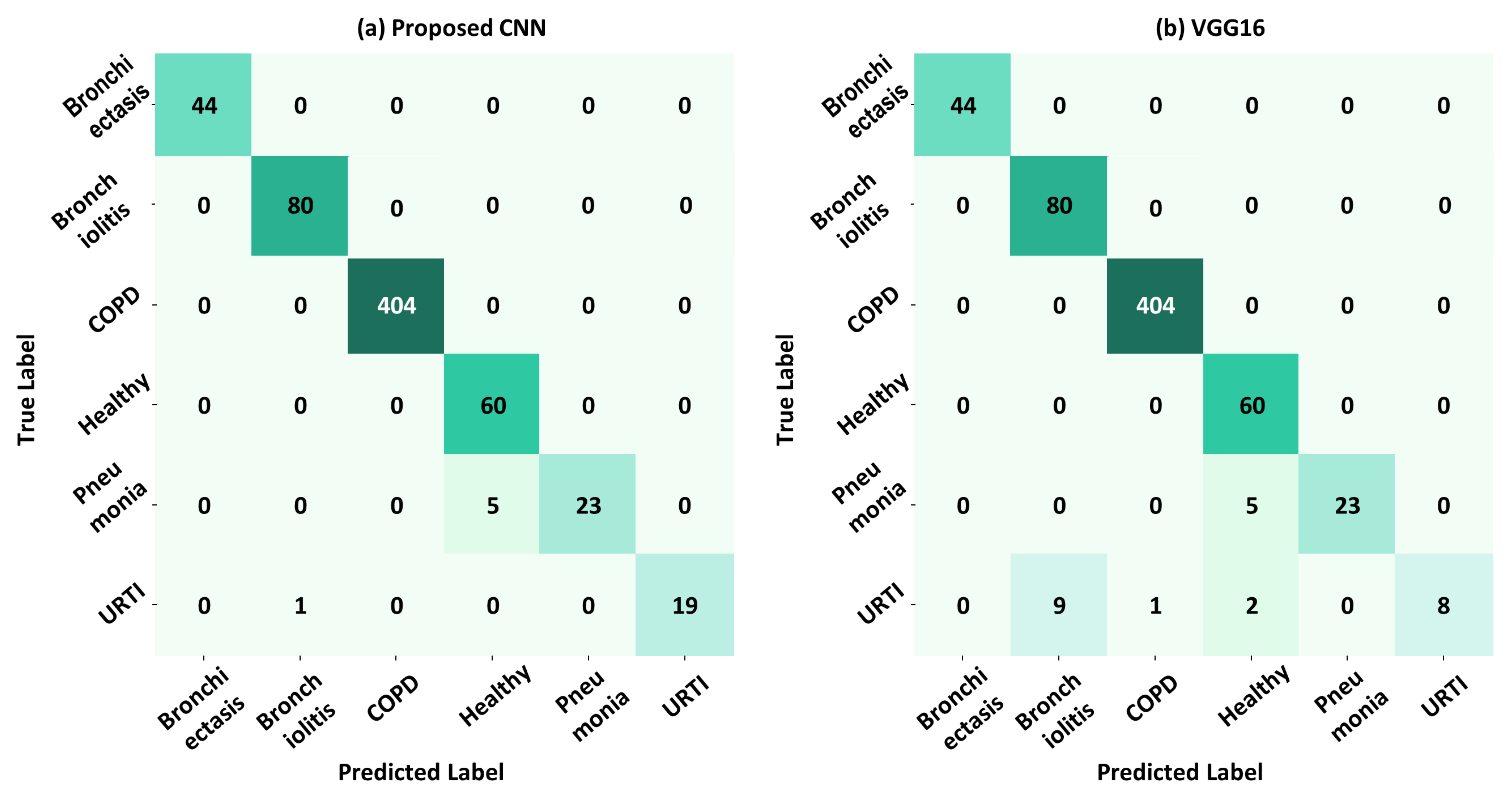}
    \caption{\textbf{Confusion matrices for the best results obtained in six-class Pathological classification. (a) Proposed CNN model with batch size 6; (b) VGG16 with batch size 6.}}
    \label{f7}
\end{figure}

For six-class Pathological classification, the proposed method involving hybrid scalogram and proposed CNN model classifier yields the best accuracy, 99.05\%, as seen in Table~\ref{tab3}. Similar to the case in the ternary chronic classification scheme, the accuracy of  VGG16 is slightly lower.  However, since the dataset gets more segregated being divided into six different disease classes, the accuracy drop is more here.  The proposed hybrid scalogram outperforms the conventional CWT scalogram with a larger margin (13.4\%-14.7\%) for both VGG16 and our proposed model. In general, the proposed method gives a better performance, which is apparent from the best confusion matrices shown in Fig.~\ref{f7}.

\subsection{Comparison with Other Works}
\subsubsection{Respiratory Disease Classification}
As discussed in section~\ref{mt2}, none of the existing works for respiratory disease classification explore the domain of patient-specific prediction. Some of the studies address the issues regarding class imbalance~\cite{vbasu2020, garcia2020vae}.Nevertheless, the extensive preprocessing, coupled with the ambiguous undersampling of the COPD disease class while oversampling all other disease classes, can complicate  the reproducibility of~\cite{vbasu2020}. Furthermore, in~\cite{garcia2020vae}, FFT is applied to the entire respiratory sound signals, whereas our work focuses on segmented breath sounds. In our work, complete patient independence has been maintained in the train and validation set, which is not possible while using the entire lung auscultation signal due to the low number of samples. Therefore, our work aims to overcome all the drawbacks present in the existing methods. A comparison among the various methods, including the Proposed method, is provided in Table~\ref{t7}.
\par 
It is observed that our proposed CNN model with the hybrid scalogram can perform on par with the existing state-of-the-art CNN and RNN models for both cases of classification while maintaining a patient independent train-validation scheme. 

\subsubsection {Computational Performance as a Lightweight Network}
A detailed comparison is presented in Table~\ref{t8} among VGG16~\cite{vgg}, our proposed CNN model, AlexNet~\cite{alexnet} and the existing state-of-the-art  lightweight models such as MobileNetV2~\cite{mobilenetv2}, NASNet~\cite{nasnet}, ShuffleNetV2~\cite{shufflenet} in terms of size, trainable parameters, the number of operations measured by multiply-add (MAdd) and accuracy on both chronic and pathological approach. In terms of accuracy, our proposed CNN model shows better results than VGG16 while requiring only 3\% of the parameters. The proposed CNN model also outperforms the contemporary lightweight models, ShuffleNet V2, MobileNet V2, and NASNet relatively by 0.16\%, 0.32\%, and 0.80\%, respectively, while obtaining better trade-off between the number of parameters, requiring significantly lower storage space and computational power. This makes our proposed lightweight model more suitable for real-time wearable devices with faster and less resource-intensive training. 

We have calculated the time required for the end-to-end classification of an auscultation sound using our framework. For this experiment, we only performed the preprocessing and inference step using all of our test data and calculated the mean and standard deviation of the required CPU time. We found that the preprocessing time for EMD+CWT is $~8s \pm 0.5s.$, and only CWT is $~7.2s\pm 0.5s$. These processes are run on a Core i7 7500 processor with a 2.70-2.90GHz speed. Time required for the classification of a scalogram using the proposed network is $ ~0.07s\pm 0.01s$, while the MobileNetV2 takes $~0.085s\pm 0.01s$. Thus, the proposed CNN is faster in classifying a sound image as compared to MobileNetV2.
\begin{table}[h!]
\centering
\caption{Comparisons among Several Models from Lightweight Perspective}
\label{t8}
\begin{tabular}{|p{0.45in}|p{0.29in}|p{0.32 in}|p{0.36 in}|p{0.30 in}|p{0.30 in}|p{0.31 in}|}
    \hline
\multirow{2}{*}{\textbf{Parameter}}&\multicolumn{6}{c|}{\textbf{Network}}\\
    \cline{2-7}
&VGG16&AlexNet&\textbf{Proposed model}&Mobile-Net (v2)&Shuffle-Net (v2)
&NASNet
\\    \hline
Size(after training)&	1.5GB&294MB&44.85MB&49MB&46.9MB&64MB\\     \hline
Trainable parameters &	138M&25.704M&3.7674M&4.2M&5.4M&4.2M\\    \hline
MAdd &154.7G&725M&371.93M&575M&564M&567M\\    \hline
Accuracy (6~classes)&	97.60\%	&98.237\%&99.05\%	&98.89\%&98.27\%&98.73\%\\    \hline
Accuracy
(3~classes)	 &97.60\%&99.519\%&99.21\%&98.72\%&99.06\%&98.42\%\\
\hline
\end{tabular}
\end{table}

\section{Conclusion}     \label{mt7}
In this work, we have proposed a lightweight CNN model to classify respiratory diseases using scalogram images of lung sounds. A hybrid approach employing both EMD and CWT is presented to generate the scalogram images. The publicly available ICBHI 2017 challenge dataset has been used for the Chronic and Pathological classification of respiratory diseases. The proposed method has provided a considerable accuracy of 99.21\% for ternary chronic classification. In pathological classification among six disease classes, an accuracy of 99.05\% is achieved. The obtained accuracies are higher than VGG16, which is a much larger network. In addition, for both cases of classifications, the proposed framework provides better or a comparable performance with respect to the existing state-of-the-art methods in terms of Precision, Recall, F1-score, Sensitivity, Specificity and ICBHI score. It is worthwhile to mention that unlike most of these methods, the classification performance of the proposed technique has been assessed, keeping the training and testing data-independent in terms of patients. The proposed classifier's computational complexity has also been compared with a number of well-known CNN models and state-of-the-art lightweight networks. It has been shown to achieve high accuracy in classification while being a lightweight deep architecture. We believe that these attributes can enable the development of the automatic classification of respiratory diseases from lung auscultations in real-world clinical applications.

\bibliographystyle{IEEEtran}
\balance
\bibliography{ref}

\end{document}